\documentclass[conference]{IEEEtran}

\usepackage{cite}

\IEEEoverridecommandlockouts

\usepackage{amsfonts}
\usepackage{graphicx,times,psfrag,xspace,color,latexsym,amsmath,amssymb}
\usepackage{wrapfig,setspace}
\usepackage{subfigure}          
\usepackage{longtable}          
\usepackage{url}                

\newcommand{\be}{\begin{eqnarray}}
\newcommand{\ee}{\end{eqnarray}}
\newcommand{\beqa}{\begin{eqnarray*}}
\newcommand{\eeqa}{\end{eqnarray*}}

 \newcommand{\oldstuff}[1]{}
\definecolor{gray}{rgb}{0.5,0.5,0.5}

\newcommand{\ie}{{\it i.e.},\xspace}
\newcommand{\eg}{{\it e.g.},\xspace}
\newcommand{\cf}{{\it cf.},\xspace}

\def \mysum {\displaystyle\sum\limits}

\renewcommand{\th}[1]{#1^{\rm th}}

\newcommand\uS{\underline{S}}
\newcommand\uD{\underline{D}}

\newcommand\up{\underline{p}}

\usepackage{pslatex} 

\begin{document}

\title{Generation bidding game  with 
flexible demand\thanks{This work
supported by NSF grant SaTC 1228717. 
An earlier conference
version of this paper, focusing on commodity flow 
experimentation, appeared in {\em Proc. USENIX Feedback Computing}, 
Philadelphia, June 2014.}
}

\author{
\centering
\IEEEauthorblockN{Y. Shan\IEEEauthorrefmark{1},
J. Raghuram\IEEEauthorrefmark{1},
G. Kesidis\IEEEauthorrefmark{1}, C. Griffin\IEEEauthorrefmark{1}, 
K. Levitt\IEEEauthorrefmark{2}, D.J. Miller\IEEEauthorrefmark{1}, J. Rowe\IEEEauthorrefmark{2}, A. Scaglione\IEEEauthorrefmark{2}}\\
\IEEEauthorblockA{\IEEEauthorrefmark{1}
CS\&E, EE, \& Math Depts  and ARL,
The Pennsylvania State University,
University Park, PA, 16802
}\\
\IEEEauthorblockA{\IEEEauthorrefmark{2}
CS and EE Depts,
The University of California at Davis,
Davis, CA, 95616
}
}

\maketitle

\section{Introduction}
Game theoretic approaches to the study of electricity markets have been explored for decades \cite{DW00,FFH04,AX05}. Recently problems associated with variations of the  optimal power flow problem with static (inelastic) demand for an electrical power grid \cite{WW96,WSM12,Chen10}, have been considered  by several authors, \eg \cite{Mitter10,Garcia11}. Indeed, demand elasticity for electricity is motivated by the onset of potentially enormous load from plug-in electric and hybrid-electric vehicles, see \eg \cite{Alizadeh13,Lu13} and the references therein.

We consider a noncooperative, iterated Bertrand-type game played by the generators based on information from the grid (independent system operator (ISO)). That is, the grid is assumed to provide sufficient information so that the generators can modify their prices to improve upon their net utility. The game is a ``discriminatory" sealed-bid auction in that the generators earn at the price they bid but in a quantity determined by the ISO \cite{DW00,FFH04}. To simplify matters herein, we do not consider strategic bidding by the generators wherein they may infer demand and/or the bidding strategies of their competitors via a probabilistic model, nor multipart bidding to account for start-up/ramp-up costs, secure contracts involving minimum and maximum supply per generator, and the like\footnote{For example, in \cite{WSM12}, an affine single-part bid and associated ``uplift" payments that are part of a joint integer-programming unit commitment and continuous-linear optimal power flow (OPF, or ``economic dispatch") problem was considered.}, nor peak-power consumption penalties \cite{Duke,FortCollins}. Also, we assume a continuously differentiable and convex cost of supply \cite{AX05}, quadratic in particular. Also, rather than demand-side bidding, we  assume a simple ``passive" linear demand response based on {\em average} cost of supply, \ie the ISO ensures that its customers pay the same rates irrespective of their location or demand volume.

Here we attempt to understand demand-response on
the wholesale (generation-level) market\footnote{Again,
simplified here by 
not considering generation constraints and costs of ramp-up in 
day-ahead provisioning, and associated reliability issues.}.
As such, we are focusing here on how a demand-response consumers (load)
can influence the wholesale generation market \cite{Chen12,VA13}.

In summary, in this paper, we are interested in studying the optimum power-flow (economic dispatch) in the presence of flexible elastic demand for a mean clearing-price based marketplace, assuming the generators are free to set their prices; however in so doing energy demand will change. 
We formulate a noncooperative 
game involving
\begin{itemize}
\item generators (suppliers, wholesalers) of a single commodity (electricity) as players, 
\item a retailer-distributor (grid, ISO) that merely delivers sufficient information to the supplier-players to act to reach a Nash equilibrium, and 
\item consumers (individual loads) who are also informed by the ISO of the current spot clearing prices for power, of course.
\end{itemize}
Consistent with an ISO, we assume that the retailer/distributor controls the conduits of supply.  

In this setting, by numerically studying the relatively 
simple benchmark IEEE 9-bus system, we found that even for a generation duopoly with neither transmission-capacity bounds nor thermal losses, and quadratic (continuous) cost of supply under DC approximation, there is a surprising plurality of Nash equilibria and surprisingly complex dynamics. Similar non-convergent/unstable phenomena were also mentioned in \cite{Chen10, Kesidis13TR, Roo12}. Although our model resembles a Bertrand game by letting customers respond to the prices offered by suppliers, the result of our experiment is not a typical Bertrand outcome, \ie the outcome of a competition wherein no competitor earns a profit. This is largely due to the strictly convex, instead of linear, form of 
generation cost and the guarantee of a minimum allocated amount of
 power generation for each generator.

In all of our numerical experiments, we found a
desirable interior Nash
equilibria with equal prices and  equal
power-generation allocations for generators.

\section{Problem set-up}\label{ideal-game-sec}

Consider a retailer with suppliers and consumers of a single commodity. As in \cite{Mitter10}, supposing that each supplier $g\in G$ sets its own price $p_g$\$ per unit commodity. 
We model aggregate consumer demand to be linear in response to clearing price, $P$\footnote{Herein, just the mean price of supply.} 
\be\label{linear-demand}
D(P) & = & (D_{\max} - D_{\min})(1 - \frac{P}{P_{\max}}) ~+~ D_{\min},
\ee
where $D_{\min}$ represents inflexible demand.

Suppose that suppliers have strictly convex cost of supply, \eg
quadratic cost\footnote{We assumed quadratic cost of supply for tractability of the duopoly studied in \cite{Kesidis13TR}. An alternative cost structure could be asymptotic to a maximum, \eg $c(0)/(s-s_{\max})$ where $c(0)$ is the cost of keeping the generator/supplier online even if zero supply is being delivered. In this paper, we do not consider ramp up/ramp-down constraints for generators/suppliers.}, so that the net utility/revenue\footnote{If the {\em net} consumer utility is collectively $V(D) - P D$, then for this linear demand-response to price (\ref{linear-demand}), the utility  is quadratic, concave and increasing, $V(D) = (P_{\max}/2) (D_{\max}^2-(D_{\max}-D)^2)/(D_{\max}-D_{\min})$ for $D_{\min}\leq D\leq D_{\max}$.\label{consumer-utility-footnote}} 
so that the $\th{g}$ supplier's net utility is
\be\label{generator-util}
u_g(\up) ~=~ p_g S_g(\up) - c_g(S_g(\up)) 
~=~ p_g S_g(\up) - a_g S_g^2(\up),
\ee
with different suppliers/generators $g$ having different $a_g$ cost parameters.


We assume that supply allocations are the result of the optimization of a supply network by a mathematical program. In electricity markets, the retailer is sometimes also the distribution system.
Consider power flow in a power system with the bus set $B$ and branch set $R$. 
\begin{itemize}
\item Let $G\subset B$ be the set of generator buses $g$, having generated power $S_g$, price per unit supply $p_g$, and minimum and maximum supply $S_g^{(\min)}$ and $S_g^{(\max)}$ respectively.
\item Let $L\subset B$ be the set of load buses $l$, having a demand $D_l$ that depends on the clearing price $P$.
\item For each bus $b\in B$, $\theta_b$ is its voltage angle.
\item Finally, let $r_{i,j}$ be the branch connecting buses $b_i$ and $b_j$, 
with reactance of the branch $x_{i,j}$, 
$P_{i,j}$ power ``flowing" from $b_j$ to $b_i$ 
(neglecting power loss on the transmission line, we get $P_{i,j}=-P_{j,i}$), and
the maximum tolerable power on the branch $c_{i,j}$. 
\end{itemize}

\subsection{Optimal power flow problem formulation}

Assuming fixed generation prices $\up$, and associated clearing price $P$, the total consumer demand is given by 
(\ref{linear-demand}).
The individual consumer demands (loads) are assumed to be some fixed proportion of the total demand, \ie $D_l(P) = \alpha_l \,D(P), ~\forall l \in L$, where $\alpha_l > 0$ and $\mysum_{\substack{l \in L}} \alpha_l = 1$. The ISO/retailer solves a constrained optimization problem in order to find $\uS = [S_1, \ldots, S_{|G|}]$, the optimum power generating assignment for the generators, and $\underline{\theta} = [\theta_1, \ldots, \theta_{|B|}]$, the voltage angles on the buses, which minimize the 
charges by the generators. 
The constrained optimization problem is thus given by: 
\begin{eqnarray}\label{OPF-problem}
&&\min_{\uS,\underline{\theta}} ~\mysum_{g \in G} p_g S_g \nonumber \\
&&\mbox{such that:} \nonumber \\
&&P_{i,j} ~ = ~ \frac{1}{x_{i,j}}(\theta_j~-~\theta_i)~~~\mbox{(DC approximation\cite{WW96})} \nonumber \\
&&S_g ~=~ \mysum_{k \in B}P_{k,g},~\forall g \in G ~~~\mbox{(power generation)} \nonumber \\
&&D_l ~=~ \mysum_{k \in B}P_{l,k}, ~\forall l \in L ~~~\mbox{(consumer demand)} \nonumber \\
&&S_g^{(\min)} \leq~ S_g \leq ~ S_g^{(\max)}, ~\forall g \in G ~~~\mbox{(supplier limits)} \nonumber \\
&&-c_{i,j} ~\leq ~ P_{i,j} \leq ~ c_{i,j}, ~\forall i,j\in B ~~~\mbox{(branch limits)} 
\nonumber \\
\end{eqnarray}
Neglecting the power loss in generation and transmission, we have $\mysum_{g\in G}S_g~=~\mysum_{l \in L}D_l$. 

In the above formulation, we used fixed upper and lower bounds on the supply allocations $S_g$. Alternatively, a quadratic penalty term in the utility function (\ref{generator-util}) can serve as a ``soft'' penalty (cost) on the supply allocation. For a positive $a_g$ and fixed supplier prices $\up$, suppose supplier $g$ wants to ensure that its utility function is never smaller than some positive value $u_g^{(\min)}$, then this imposes lower and upper bounds on its supply allocation, given by
\[
\frac{1}{2 a_g}(p_g - \sqrt{p_g^2 - 4 a_g u_g^{(\min)}}) \leq S_g \leq
 \frac{1}{2 a_g}(p_g + \sqrt{p_g^2 - 4 a_g u_g^{(\min)}}),
\]
provided the supplier price satisfies the condition $p_g \geq 2\sqrt{a_g u_g^{(\min)}}$. If $u_g^{\min} = 0$, then we observe that as $a_g$ is made larger, \ie as the cost of supply allocation increases, the maximum supply allocation (or capacity) decreases, and vice-versa. 
   
In practice for power-transmission circuits, thermal losses may determine edge (transmission line) capacities and costs, the latter typically in a power-flow dependent fashion, \eg ``$I^2R$" losses (Sec. 3.1 of \cite{WW96}). In order to focus on the bidding behavior among the generators, we neglect the power loss on the transmission lines; hence the cost in power transmission is also neglected.

\subsection{Set-up of suppliers' iterative game on a platform of demand response}\label{sec:iterative_game}
We neither assume that each supplier's cost of production is known to the retailer (\ie the $a_g$ terms), nor, equivalently, that the retailer chooses its allocations to the suppliers based on this (as in \cite{Mitter10}). In the following, denote  as $\uS(\uD(P), \up)$ the solution of the above optimal power flow allocation problem to determine supply allocations for {\em fixed} demands (which are based on the clearing price $P$) $\uD(P) = \{D_j(P) \,|\, j \in J\}$ and fixed supplier prices $\up = \{p_g \,|\, g\in G\}$. We propose the following iterative supplier game wherein, for fixed supplier prices, the clearing price and the consumer demands are adjusted iteratively until they converge to a fixed point. Then each supplier $g \in G$ adjusts its price $p_g$, given the current price for all other suppliers $\up_{-g}$, such that its utility function $u_g(p_g, \up_{-g})$ is increased. Given initial prices set by the suppliers $\up$, the iterative supplier game proceeds as follows:
\begin{enumerate}
\item The retailer/ISO sets an {\em initial} mean price of supply (clearing price charged to all consumers), $P$, say just as the mean of the initial supplier prices, $p_g, ~\forall g \in G$.
\item Determine the price-dependent consumer demands $\uD(P)$, where $D_j(P) = \alpha_j \,D(P), ~\forall j \in J$.
\item Retailer solves the economic dispatch optimal power flow allocation problem $\uS(\uD(P), \up)$ given fixed demands $\uD$ and generation/supply costs $\up$.
\item Retailer computes a new mean (clearing) price of supply, $P ~=~ \mysum_{g \in G} S_g \,p_g/\mysum_{g \in G} S_g$.
\item If the change in clearing price $P$ is significant (larger than some threshold), then go back to Step 2; else continue to Step 6. 
\item For the current set of supplier prices, consistent supply allocations, consumer demands, and clearing price have been found. Now each supplier sets a new price of supply such that there is an increase in its utility function using one of the following two approaches: \\
(i) \emph{Best-response play action}:
Each supplier $g$ sets a new price of supply based on (an estimate 
of)\footnote{Since a closed-form solution to the objective is not found, 
we use a small positive step-size to directly search for optimal 
prices within their defined range.}
\be\label{best-response}\label{best_resp_eq}
\arg\max_{p_g} ~p_g \,S_g(p_g; \up_{-g}) ~-~ c_g(S_g(p_g; \up_{-g})),
\ee
where $c_g(x)$ is the cost of supply (assumed $=a_g x^2$ above). \\
(ii) \emph{Better-response play action}:
Each supplier $g$ calculates approximate left and right partial derivatives of its utility function with respect to its price $p_g$, \ie
\begin{eqnarray}
&&\Delta u_g^+ ~=~ \frac{u_g(p_g + \epsilon,\, \up_{-g}) - u_g(p_g,\, \up_{-g})}{\epsilon} \nonumber \\
&&\Delta u_g^- ~=~ \frac{u_g(p_g,\, \up_{-g}) - u_g(p_g - \epsilon,\, \up_{-g})}{\epsilon}, \nonumber
\end{eqnarray}
where $\epsilon \searrow 0$\footnote{We chose a value of $\epsilon = 10^{-6}$.}. If the left and right derivatives have different signs (a non-differentiable point), then there are two possibilities. If $\Delta u_g^- > 0$ and $\Delta u_g^+ < 0$, the current price $p_g$ is a local maximum and there is no need to change $p_g$. If $\Delta u_g^- < 0$ and $\Delta u_g^+ > 0$, the current price is a local minimum. In this case, we increase $p_g$ by a small value $\zeta$ if $|\Delta u_g^+| > |\Delta u_g^-|$; otherwise we decrease $p_g$ by $\zeta$. In case the derivatives have the same sign (may still be a non-differentiable point), we increase $p_g$ by $\zeta$ if both derivatives are positive and decrease $p_g$ by $\zeta$ if both derivatives are negative. The step $\zeta$ should increase the price by a small value such that there is an increase in the value of the utility function. It should not make large changes to the price like the best-response play action\footnote{We chose $\zeta$ as follows. Starting with a small trial value of $\zeta = 0.005\,p_g$, if $u_g(p_g + \zeta,\, \up_{-g}) > u_g(p_g,\, \up_{-g})$ we accept the value of $\zeta$, else $\zeta$ is decreased by a factor of $2$ iteratively until $u_g(p_g + \zeta,\, \up_{-g}) > u_g(p_g,\, \up_{-g})$.\label{fnt:btt_step}}. 
\item Exit if there is no change in the supplier prices (\ie if an equilibrium set of prices is obtained); Else go back to Step 1. 
\end{enumerate}

Even for simple power circuits,
the best-response iterated play may lead to convergence problems including
limit-cycle behavior (\cf Section \ref{numerical-sec}). 
Alternatively, the suppliers could play the iterated better-response 
non-cooperative game, possibly with more reliable convergence 
properties\cite{ToN05,Shamma05}. 

Given global knowledge of the retailer's supply conduits\footnote{Again, the motivating example here is a power system retailer/ISO  which owns and operates the grid connecting generators/suppliers to loads/consumers.},
each supplier can compute its ``best-response" prices in Step 6 leading to a Nash equilibrium. Alternatively, the retailer may not explicitly divulge its system state (just as the cost of supply is not known to the retailer in our set-up) and compute the revenue function $f_g(p_g) = p_g \,S_g(p_g; \up_{-g})$ for each supplier $g\in G$, again assuming $\up_{-g}$ fixed from the previous iteration.

Note how this algorithm depends on forecasts of demand for the upcoming epoch (which needs to  be long enough to accommodate the ramp-up/down delays of the generators). Day-ahead forecasts \cite{Rahimi10} could be used to inform the initial prices set by the generators.

\section{Numerical study}\label{numerical-sec}
\begin{figure}[ht!]
\includegraphics[trim=120 0 120 0,clip,scale=0.35]{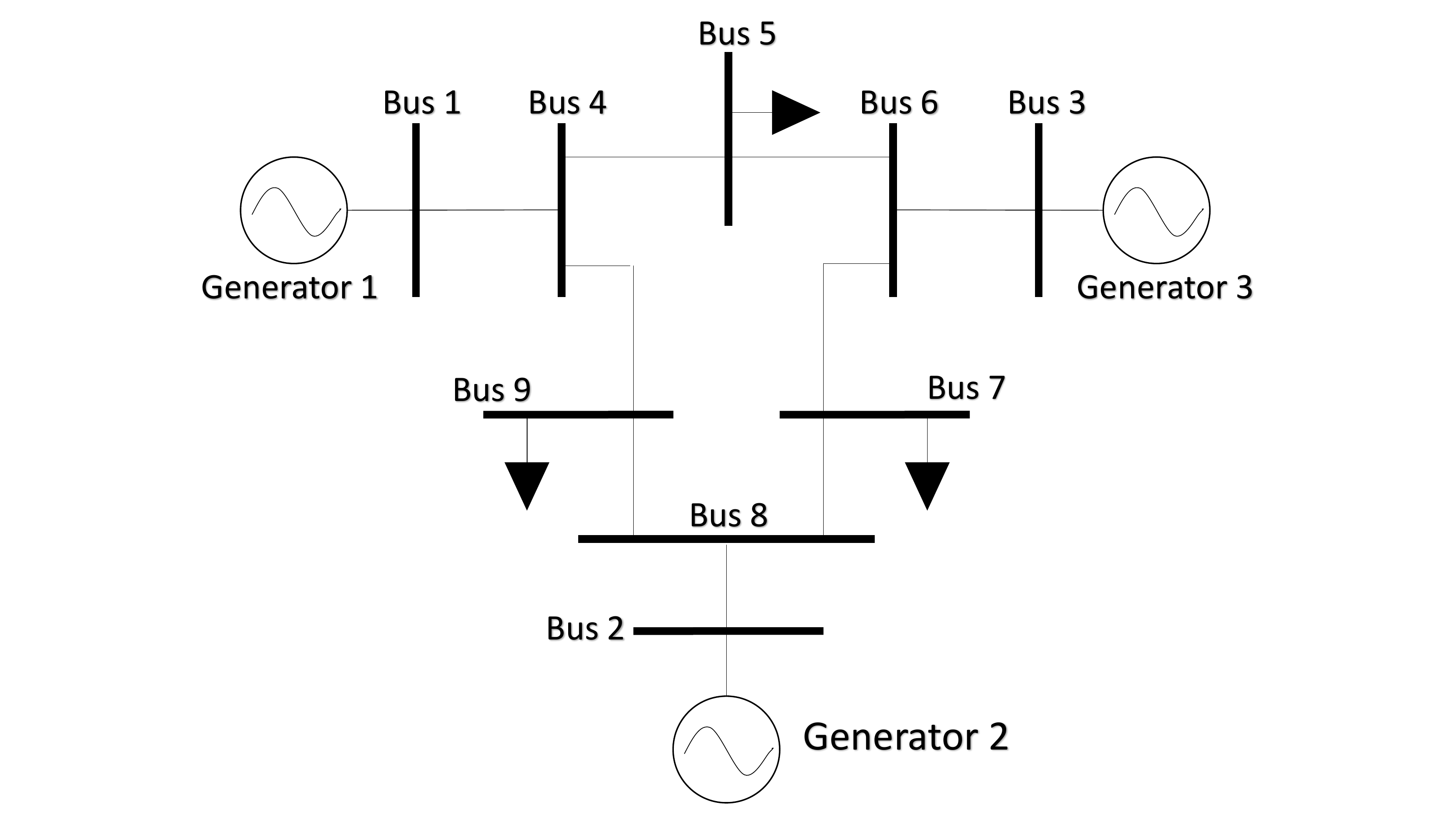}
\caption{IEEE 9-bus system
used in our study
with three generators and three loads}\label{fig:9bus}
\end{figure}

We study the iterative generation game described in section \ref{sec:iterative_game} with a 9-bus power system that has three generators and three loads as shown in Fig. \ref{fig:9bus}. 
We consider the scenario where cost of delivering power is zero, and the branch capacities are set to fixed values. 

For the benchmark IEEE 9-bus system\cite{MATPW}, the maximum and minimum power generated by the three generators are set to $[250, 300, 270]$MW and $[10, 10, 10]$MW, respectively. The constants in $D(P)$, our model for the total consumer demand, were chosen as the possible maximum/minimum power provided by the generators, that is, $D_{\max}=770$MW, $D_{min}=30$MW (note that, although generator 2 can generate 300MW, the capacity of the branch connecting bus 2 and bus 8 is only 250MW, therefore generator 2 can only provide 250MW at most). The total consumer demand $D(P)$ is assumed to be proportionally divided among the individual loads, \ie $\alpha_l = \frac{1}{|L|}, ~\forall l \in L$. The maximum clearing price, $P_{max}$, is set to 5; for clearing prices $P > P_{\max}$, the flexible demand is $0$. The reactance of the transmission line is based on the data from \cite{JHC}. The constants ($a_g$) in the utility function (\ref{generator-util}) of generators are set to $0.009,~ 0.01$, and $0.018$ respectively. 

\subsubsection{Best response game with three players} 
Assume all three generators participate. By choosing different starting points, we have both convergent and non-convergent trajectories, as shown below. 

\begin{figure}[ht!]
\includegraphics[trim=25 90 25 90,clip,scale=0.43]{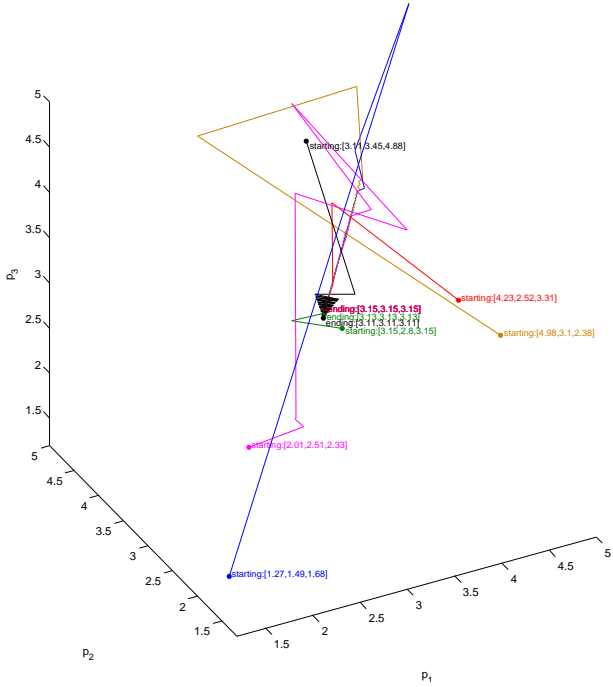}
\caption{The three-player best-response trajectories which start from different initial prices and end up with equilibrium points.}
\label{fig:c_traj}
\end{figure}

In Fig. \ref{fig:c_traj}, these trajectories will finally converge to definite points. For most initial generator prices, the limit pricing point is $[3.15, 3.15, 3.15]$, however there are several exceptions. For example, trajectories starting from $[3.11, 3.45, 4.88]$ and $[3.15,2.8,3.15]$ converge to $[3.11, 3.11, 3.11]$ and $[3.13, 3.13, 3.13]$, respectively. We found that when the prices fall at the points along the line from $[2.61,2.61,2.61]$ to $[3.15,3.15,3.15]$, all the generator will keep their prices unchanged, which means, once a trajectory intersects with this line, it terminates at the intersection point (convergence occurs). 
At these equal-prices equilibrium points, 
there are equal power allocations $S$ to the generators.

\begin{figure}[ht!]
\includegraphics[trim=20 90 25 150,clip,scale=0.43]{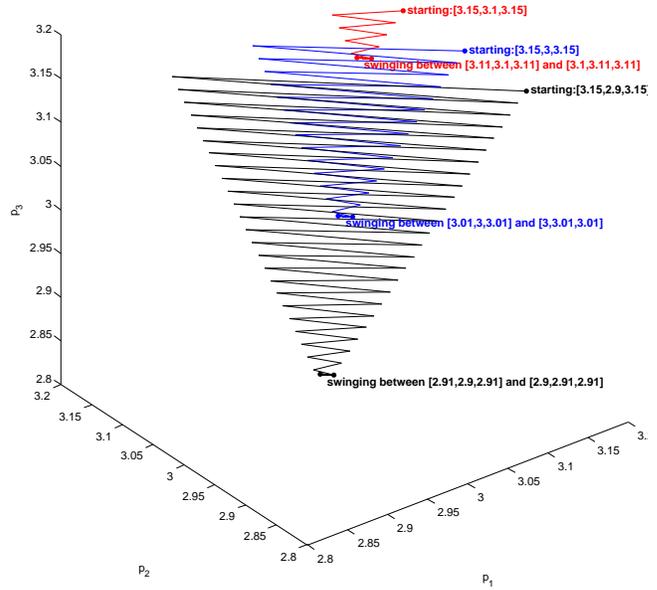}
\caption{The three-player best-response trajectories which end up oscillating between two points.}
\label{fig:nc_traj}
\end{figure}

On the other hand, we also found some best-response trajectories which do not converge, as shown in Fig. \ref{fig:nc_traj}, because of the finite step-size 
when searching for optimal prices, given the current prices of other generators (Step 6 in \cf Section \ref{sec:iterative_game}). In our 
numerical experiments, we set the step-size to 0.01.

\subsubsection{Best response game with two players}
By fixing the price of generator 3 to $5$, we can focus on the best-response game between generators 1 and 2, and plot the trajectory in a two-dimensional plane. Similar to three-player best-response, we also found both convergent and non-convergent trajectories. 
 
\begin{figure}[ht!]
\includegraphics[trim=0 100 0 100,clip,scale=0.4]{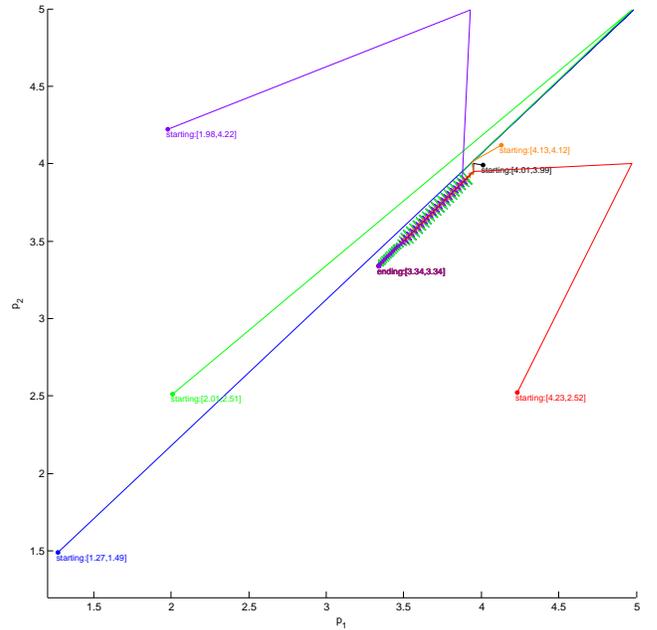}
\caption{The two-player best-response trajectories which start from different points and converge to the same point.}
\label{fig:c_traj2}
\end{figure}

However, our numerical results presented in Fig. \ref{fig:c_traj2} show that, 
unlike the games with three players, the best-response games with two competitors, if convergent, will converge to the point $(3.34,3.34, 3.34)$).

\begin{figure}[ht!]
\includegraphics[trim=0 100 0 100,clip,scale=0.4]{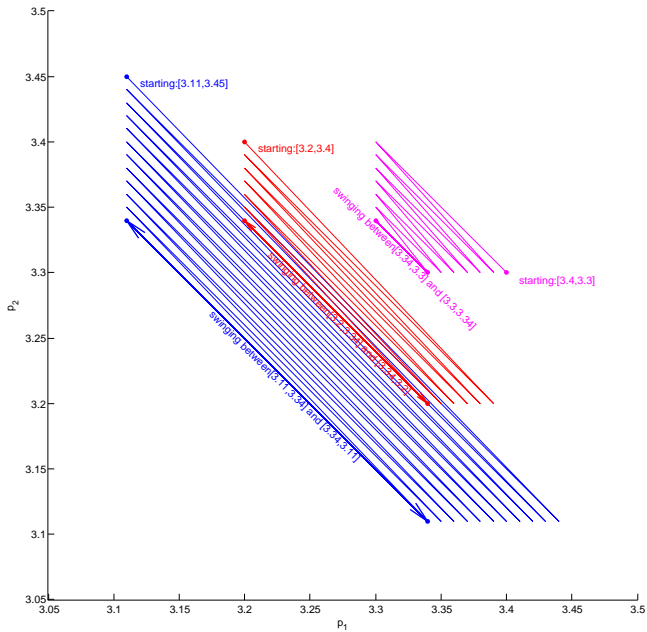}
\caption{The two-player best-response trajectories which start from different points and end up oscillating between two points.}
\label{fig:nc_traj2}
\end{figure}

Moreover, non-convergent phenomena in the two-player best-response game 
are more manifest than in the three-player game. 
As demonstrated in Fig. \ref{fig:nc_traj2}, the amplitude of oscillation, compared with 2 players, is much larger. Also, unlike the non-convergent cases in the three-player game, decreasing the step length will not shorten the oscillation amplitude.

\subsubsection{Better response game with two players}
\begin{figure}[ht!]
\includegraphics[trim=0 100 0 100,clip,scale=0.4]{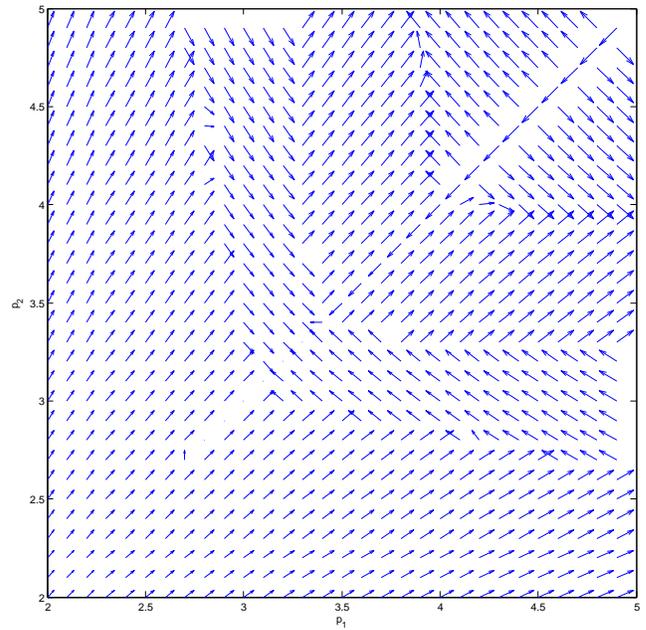}
\caption{Quiver plot of the better response dynamics with price of generator 3 set to 5.}
\label{fig:quiver_plot}
\end{figure}

 We simulate two-player better-response game, an alternative approach, which avoids the oscillatory behavior of the best-response game. The quiver plot shown in Fig. \ref{fig:quiver_plot} can visualize the dynamics of better-response iterated play microscopically. Note that we have set an upper bound price to 5,
otherwise prices will diverge from an infinitely large set of initial prices. 
This is because we set a minimum power allocation 
for each generator in our optimal power flow calculation, \ie the generators always have some power to generate, no matter how high the prices are. 
However, such minimum allocated amounts of generation power typically
are not guaranteed irrespective of price in practice.

\begin{figure}[ht!]
\includegraphics[trim=0 100 0 100,clip,scale=0.39]{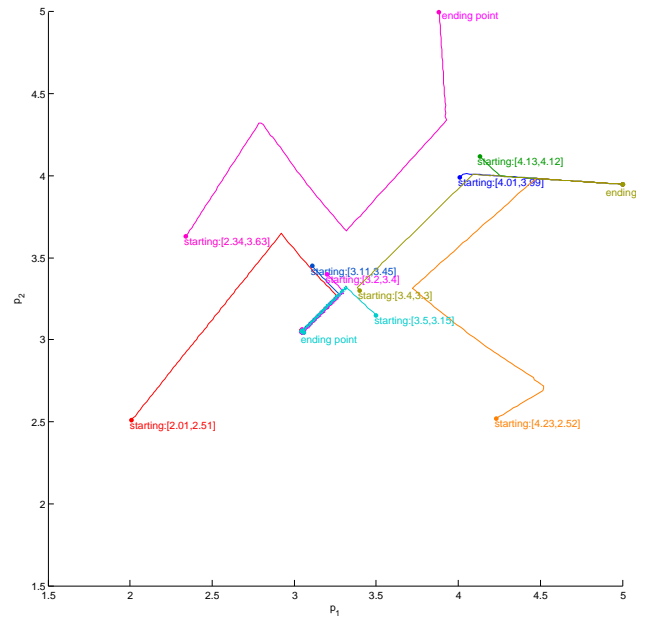}
\caption{The two-player better-response trajectories starting from different points, the mass around point $(3.05,3.05)$ can be approximated to a point if the step length is small enough.}
\label{fig:btt_traj}
\end{figure}

Based on the quiver plot shown in Fig. \ref{fig:quiver_plot}, we conjecture that trajectories starting from some area (most likely the lower left area) will converge to an interior NEP, while those starting from the other area will converge to the boundary (points where either $p_1$ or $p_2$ is 5), as demonstrated in Fig. \ref{fig:btt_traj}. Note that, in our experiment, the better-response trajectories with an interior NEP actually do not converge, but arrive at
small limit cycles because of finite step-size for search,
recall footnote \ref{fnt:btt_step}.

\bibliographystyle{plain}

\end{document}